\documentclass[twocolumn,aps,prb,longbibliography,floatfix,10pt]{revtex4-2}

\usepackage{natbib}
\usepackage{times}
\usepackage{graphicx}
\PassOptionsToPackage{caption=false}{subfig}
\usepackage{subfig}
\usepackage[percent]{overpic}
\usepackage{adjustbox}
\usepackage{amsfonts}
\usepackage{amsmath,amsthm,amssymb,mathrsfs}
\usepackage{dsfont}
\usepackage{xcolor}
\usepackage{microtype}
\usepackage{cancel}
\usepackage{bm}
\usepackage{siunitx}
\usepackage[colorlinks=true, allcolors=black, citecolor=blue, linkcolor=blue, urlcolor=blue]{hyperref}
\usepackage[integrals]{wasysym}

\usepackage[capitalize]{cleveref}

\begin{document}

\title{Eddy currents and current reversal in curved magnetic thin-film Josephson junctions}

\author{Einar Skoglund}
\author{Maxim A. Tjøtta}
\author{Sol H. Jacobsen}
\thanks{Corresponding author: sol.jacobsen@ntnu.no}
\affiliation{Center for Quantum Spintronics, Department of Physics, Norwegian University of Science and Technology NTNU, NO-7491 Trondheim, Norway}

\begin{abstract}
    Real-space geometric curvature in magnetic thin films introduces a controllable mechanism for tailoring the pathways of superconducting steady-state Josephson currents via the proximity effect. We present a generalized Green's function method for calculating diffusive transport in arbitrarily curved surfaces, and show how the competing mechanisms of curvature and distance regulate conversion between different superconducting pairings in a proximity-coupled ferromagnet. We show how this dictates the distribution of current density, with the possibility of curvature-controlled current density manipulation, induced eddy currents and current reversal. 
\end{abstract}

\maketitle

Curvilinear magnetism has proven to be exceptionally versatile in its field of application, from probing fundamental material properties to flexible monitoring of biological activity
\cite{Sheka2015,Streubel2021,Gentile2022, MakarovBook2022}. As material synthesis and techniques for creating and manipulating 3D curvilinear structures have become reliable also down to the nanometer range \cite{Makarov2022}, there is growing excitement about the possibility of functionalizing curvature in computation and quantum sensing. For example, compressive buckling can give versatile control of intricately curved arrays of mesoscopic wires \cite{Xu2015}, and glancing angle deposition can produce magnetic helices of a few tens of nanometers \cite{Gibbs2014}. More recently, three-dimensional chiral magnetic ribbons have been shown experimentally to enable current-induced motion of chiral domain walls, raising the prospect of 3D racetrack memories and logic devices \cite{Parkin2025}. 

Josephson junctions are nanoscale heterostructures comprised of two superconductors separated by a non-superconducting barrier, such as an insulator, normal metal or ferromagnet, through which a steady-state current can flow when there is a phase difference between the superconductors \cite{Josephson1962}. Josephson junctions are the fundamental building block of quantum devices ranging from the most sensitive superconducting quantum interference device magnetometers to transmon qubits for quantum computing \cite{kim2025,Fagaly2006}, and for this reason their behaviour in novel circuitry is still intensively studied. In superconducting spintronics, which aims to generate spin-polarized supercurrents to combine the functionalities of resistance-free currents and switchable spin states \cite{Linder2015, Eschrig2015}, a component of conventional singlet superconducting correlations can be converted into spin-polarized triplet pairs by introducing magnetic inhomogeneity between the superconductors \cite{Bergeret2001,Kadigrobov_2001}, typically via magnetic multilayers \cite{Khaire2010}, a conical field magnet \cite{Robinson2010}, or a homogeneous magnet with intrinsic or proximitized spin-orbit coupling \cite{Bergeret2013,Jacobsen2015b}. Since real-space curvature engenders a magnetically inhomogeneous ground state for many magnetic materials, curvilinear magnetism has recently been suggested as an additional way to generate and control spin-polarized superconducting correlations via the proximity effect \cite{Salamone2021,Salamone2022}. In one dimensional curved ferromagnet junctions, this has been shown theoretically to lead to control of both the superconducting transition \cite{Salamone2022, Salamone2024spin_valves}, and cause current reversal \cite{Salamone2021, Skarpeid2024}. 

In ferromagnetic Josephson junctions, it is well established that the length of the interstitial magnet governs the ground state direction of Josephson current due to the different length-dependent phase accruals of the majority and minority spin bands \cite{Ryazanov2001, Ryazonov2004, Buzdin2005, Birge2024}. In junctions with spin-orbit coupling, the current direction can also be controlled by varying the relative angle between the exchange field and spin-orbit vectors \cite{Jacobsen2015b, Arjoranta2016, Francica2020}. In curved junctions with multiple interfaces, the ground state depends not only on the length, but also on the magnetic chirality at each interface \cite{Skarpeid2024}. In this article, we will show how the current densities of the ground state are distributed and controlled in arbitrarily curved ferromagnetic thin film Josephson junctions, where the length effect competes with magnetic chirality in two transport degrees of freedom. We will use this to isolate a purely geometrically driven current reversal.

To model Josephson currents in magnetic thin films, we will use the quasiclassical Green's function theory for diffusive transport of superconducting correlations \cite{Belzig1999,Chandrasekhar2004}, via the Usadel equation \cite{Usadel1970} and Kupriyanov-Lukichev boundary conditions \cite{KuprianovLukichev1988}. This describes materials and heterostructures with significant impurities or interface roughness, and so covers a large class of experimentally accessible systems. While typically used in 1D, the Usadel method has been extended to be solved numerically for flat thin films in 2- and 3D via a finite element method \cite{Amundsen2016}. Here, we present a generalized framework for diffusive Green's function transport along any arbitrarily curved surface. We demonstrate the method for an annular sector and helicoid, which are close to the experimentally implemented twisted nanoribbons \cite{Parkin2025} and free-standing Josephson membranes \cite{Yan2024,Li2026_MembraneReview}. In addition to the full numerical solution, these simple geometries also permit analytic solution in the weak proximity limit, through which we gain insight into the dominant physical mechanisms. 

\paragraph*{Theory.}
We parameterize a curved ferromagnetic film according to a general, curved $2$D surface, denoted by $\bm{\mathcal{S}}(u,v)$. The $3$D coordinate vector $\bm{R}$ is then $\bm{R}(u,v,n)=\bm{\mathcal{S}}(u,v)+n\hat{\mathcal{N}}(u,v)$, where $\hat{\mathcal{N}}(u,v)$ denotes the normalized normal vector to the parameterized surface $\bm{\mathcal{S}}(u,v)$ \cite{daCosta1981, Pressley2010}. Such a general parameterization of $3$D space necessitates the use of tensor notation. Accordingly, we choose the basis of our coordinate system to consist of covariant and contravariant basis vectors. The covariant basis vectors are defined from the derivatives of the $3$D coordinate vector, $\bm{e}_i=\partial_i\bm{R}(u,v,n)$, $i=\{u,v,n\}$, and the contravariant basis vectors are found by raising the index of the covariant basis vectors using the inverse metric tensor, $\bm{e}^i=\mathcal{G}^{ij}\bm{e}_j$ \cite{neuenschwander2014tensor}. The metric tensor is defined as $\mathcal{G}_{ij}=\bm{e}_i\cdot\bm{e}_j$, and its inverse as $\mathcal{G}^{ij}=\bm{e}^i\cdot\bm{e}^j$ (see Appendix~\ref{Appendix:A} for details). The covariant and contravariant basis vectors thus form a dual vector space, $\bm{e}^i\cdot\bm{e}_j=\delta^i_j$, ensuring orthonormality. A general vector $\bm{A}$ can thus be expressed in both the covariant basis, $\bm{A}=A^i\bm{e}_i$, and the contravariant basis, $\bm{A}=A_i\bm{e}^i$, such that inner products take the form $\bm{A}\cdot \bm{B}=A^iB_i=\mathcal{G}^{ij}A_iB_j$. Due to the spatial dependence of the basis vectors, derivatives of vector quantities produce additional terms represented by Christoffel symbols, defined by $\partial_i\bm{e}^j=-\Gamma^j_{ik}\bm{e}^k$ or (equivalently) $\partial_i\bm{e}_j=\Gamma^k_{ij}\bm{e}_k$, where $\partial_i$ is the covariant derivative and we are summing over repeated indices.

In diffusive equilibrium, we can generalize the Usadel equation \cite{Usadel1970} for the retarded quasiclassical Green's function $\hat{g}^R$ for an arbitrarily chosen coordinate system by replacing all derivatives with their space-gauge covariant derivative counterparts, and writing scalar products using tensor notation \cite{Salamone2022, Svendsen2021}. The geometrically invariant Usadel equation thus becomes
\begin{equation}\label{eq:Usadel}
    iD_F\mathcal{G}^{ij} \Big(\partial_i\left(\hat{g}^R\partial_j\hat{g}^R\right)- \Gamma^k_{ij}\hat{g}^R\partial_k\hat{g}^R\Big)=\left[\epsilon\hat{\tau}_3+\hat{M},\hat{g}^R\right],
\end{equation}
where $D_F$ is the diffusion constant, $\epsilon$ is the quasiparticle energy, $\hat\tau_3=\text{diag}(1,1,-1,-1)$, and $\hat M=\bm h\cdot \text{diag}(\bm \sigma, \bm \sigma^*)$ is the magnetization, described by the exchange field $\bm h$ and Pauli vector $\bm \sigma=\sigma_i\bm{e}^i$. The sum over the indices in Eq.~\eqref{eq:Usadel} should be taken over the relevant directions of the system, such that for a thin film we sum over $u,v$.
For a system consisting of multiple materials, the Kuprianov--Lukichev boundary conditions at the interface between a material $1$ and its adjacent material $2$ are given by \cite{KuprianovLukichev1988}
\begin{equation}\label{eq:KL_bc}
\mathcal{G}^{ij}n_i\cdot \hat g_1^R\partial_j \hat g_1^R
=\frac{1}{L_1\zeta_{1}}\left[\hat g_1^R,\hat g_{2}^R\right],
\end{equation}
where $\bm n=n_i\bm{e}^i$ is the unit normal vector to the interface, directed outward from material $1$ towards material $2$, $L_1$ is the characteristic length of material $1$, and $\zeta_{1}=R_b/R_1$ is the ratio of the interface resistance $R_b$ to the bulk resistance $R_1$ of material $1$. From analytic work in one dimension \cite{Salamone2022}, it is clear that the addition of curvature effects in the Usadel equation~\eqref{eq:Usadel} induces rotation between different species of triplet Cooper pairs, which is of course also the case for thin films (see Appendix~\ref{Appendix:WPL} for details). Since spin-polarized and non-spin-polarized triplets couple differently to an effective field inhomogeneity, we can expect to see signatures of the different triplet species in observables that scale with the inhomogeneity.

By solving the Usadel equation for the quasiclassical Green's function, we can calculate observable quantities. Here, we will consider the charge current density $\bm J=J_i\bm{e}^i$, where
\begin{equation}\label{Eq:ChargeCurrentDensity}
    J_i=J_0 \int_0^\infty d\epsilon\:\text{Re}\:\text{Tr}\left\{\hat\tau_3\hat g^R \partial_i \hat g^R\right\}\tanh\frac{\epsilon}{2k_B T},
\end{equation}
with $J_0=N_0 eD_F\Delta/16\xi$, where $N_0$ is the density of states at the Fermi energy, $e$ is the elementary electronic charge, $k_B$ is the Boltzmann constant, $T$ is the temperature, $\Delta$ is the magnitude of the superconducting gap, and $\xi$ is the superconducting coherence length. In curved thin films, we will get a multi-component interconvergence between different triplet species. This manifests in the charge current density, which we can see by decomposing it into its constituent singlet and triplet expressions, where the triplet component has both a curvature-dependent and independent part.

To do this, we write the Green's function in terms of its normal $(g)$ and anomalous $(f)$ components
\begin{equation}
    \hat g^R = \begin{pmatrix}
        g & f \\
        -\tilde f & -\tilde g
    \end{pmatrix},
\end{equation}
where the normalization condition $(\hat{g}^R)^2=1$ implies the constraints $(gg-f\tilde{f})=1$ and $(gf-f\tilde{g})=0$. The tilde conjugation is obtained by taking the complex conjugate and reversing the sign for the energy $\tilde{g}(x,+\epsilon)\equiv g^*(x,-\epsilon)$ \cite{Jacobsen2015b}. The charge current density then depends only on the anomalous Green's function $f$, which can be split into a singlet component $f^{(s)}$ and a triplet vector component $d_j$ as $f=(f^{(s)}+d^j \sigma_j)i\sigma_y$.
The charge current density can then be written as the sum of the singlet $(s)$ and triplet $(t)$ components
$J_i/J_0 = J_i^{(s)} + J_i^{(t)},$
where
\begin{equation}
    J_i^{(s)}=-8\int_0^\infty d\epsilon\:\text{Re}\left(\tilde f^{(s)}\partial_i f^{(s)} -f^{(s)}\partial_i\tilde f^{(s)}\right)\tanh\frac{\epsilon}{2k_B T},
\end{equation}
and we have split $J_i^{(t)} = J_i^{(t_0)}+J_i^{(\Gamma)}$ into a curvature independent part $J_i^{(t_0)}$ and a curvature dependent part $J_i^{(\Gamma)}$, with
\begin{align}\label{eq:current_triplet_0}
    J^{(t_0)}_i
    &= 8\mathcal G^{jk}\int_0^\infty d\epsilon\:\text{Re}\left(\tilde d_k \partial_i d_j - d_k \partial_i \tilde d_j\right)\tanh\frac{\epsilon}{2k_B T} ,
    \\
    J^{(\Gamma)}_i 
    &= 8\mathcal G^{jk}\:\Gamma^l_{ij}\int_0^\infty d\epsilon\:\text{Re}\left(d_k\tilde d_l-\tilde d_k d_l\right)\tanh\frac{\epsilon}{2k_B T}.\label{eq:current_triplet_Christoffel}
\end{align}
It is clear from Eq.~\eqref{eq:current_triplet_Christoffel} that non-zero Christoffel symbols will yield an additional component to the current density that arises due to the interconversion between different triplet components, and scales with the factors in the Christoffel symbols, which will depend on the specific geometry in question. We note that all triplet components, both with and without finite spin projection, are included in the current density expressions of Eqs.~\labelcref{eq:current_triplet_0,eq:current_triplet_Christoffel}. In the absence of curvature, $\kappa\rightarrow 0$, the spin-polarized triplet component vanishes, such that we can identify Eq.~\eqref{eq:current_triplet_0} as the unequal-spin triplet component in this limit.

\paragraph*{Example: helicoidal strip.}
To see this framework in practice, we consider a helicoidal strip. In analogy with a 1D helical wire characterized by its curvature $\kappa$ and torsion $\tau$ \cite{Salamone2024spin_valves}, we parametrize the surface as \cite{Krivoshapko2015, ortix2015quantum}
\begin{equation} \label{eq:helicoid_surface}
    \bm{\mathcal{S}}(u,v)=\left(\frac{\kappa}{\omega^2} + v\right)\big(\cos(\omega u)\:\hat{\bm{x}}+ \sin(\omega u)\:\hat{\bm{y}}\big)+\frac{\tau u}{\omega}\:\hat{\bm{z}},
\end{equation}
where $\omega=\sqrt{\kappa^2+\tau^2}$, $u\in[0,L]$, and $v\in[-W/2,W/2]$. Here, $L$ is the length of the central helix, $W$ is the strip width, and $\kappa$ and $\tau$ denote the curvature and torsion of the central helix, respectively. Taking the limit $n\rightarrow 0$, which corresponds to a thin film surface, the covariant basis vectors are
\begin{align}
    \bm{e}_u&=\left(\frac{\kappa}{\omega}+\omega v\right)\big(-\sin(\omega u)\:\hat{\bm{x}}+\cos(\omega u) \:\hat{\bm{y}}\big)+\frac{\tau}{\omega}\:\hat{\bm{z}}, \\ \bm{e}_v&=\cos(\omega u)\:\hat{\bm{x}}+\sin(\omega u) \:\hat{\bm{y}}, \\ \bm{e}_n&=\hat{\mathcal{N}}(u,v)=\bm{e}_u\times \bm{e}_v/|\bm{e}_u\times \bm{e}_v|
\end{align}
and the metric is $\mathcal{G}=\text{diag}(\eta(v),1,1)$, where $\eta(v)=1+2\kappa v+\omega^2v^2$. The non-zero Christoffel symbols for this example are $\Gamma^v_{uu}=-\kappa-\omega^2v$ and $\Gamma^u_{uv}=\eta^{-1}(\kappa+\omega^2v)$. Inserting this into the Usadel equation~\eqref{eq:Usadel}, we get 
\begin{equation}\label{eq:Usadel_helicoid}
    \begin{split}
        iD_F &\Big(\eta^{-1}\partial_u\left(\hat{g}^R\partial_u\hat{g}^R\right) +\partial_v\left(\hat{g}^R\partial_v\hat{g}^R\right)\\&+ \eta^{-1}(\kappa +\omega^2v)\hat{g}^R\partial_v\hat{g}^R\Big)=\left[\epsilon\hat{\tau}_3+\hat{M},\hat{g}^R\right].
    \end{split}
\end{equation}
In the weak proximity limit, the anomalous Green's function $f\ll1$, $\gamma=f/2$ and $N=1$. The equations of motion for the triplet components in this limit then become (see Appendix~\ref{Appendix:WPL} for details)
\begin{align}
    \begin{split}
        D_F\big(\eta^{-1}\partial_u^2d_u+\partial_v^2&d_u +\Gamma^u_{uv}(2\partial_ud_v-\partial_vd_u)\big)\\&=\eta^{-2}\tau^2d_u-2i(Ed_u+f_0h_u),
    \end{split} \label{eq:wpl_helicoid_u} \\
    \begin{split}
        D_F\big(\eta^{-1}\partial_u^2d_v+\partial_v^2&d_v -\Gamma^u_{uv}(2\eta^{-1}\partial_ud_u- \partial_vd_v)\big)\\&=(\Gamma^u_{uv})^2d_v-2i(Ed_v+f_0h_v),
    \end{split} \label{eq:wpl_helicoid_v}
\end{align}
where we have used $-\eta^{-1}\Gamma^v_{uu}=\Gamma^u_{uv}$. In two dimensions, the equation for $d_n$ is decoupled from the rest and so does not feature.
Inserting the Christoffel symbols for the helicoid strip into Eq.~\eqref{eq:current_triplet_Christoffel}, we get $J_v^{(\Gamma)}=0$ and 
\begin{equation} \label{eq:current_christoffel_helicoid}
    J^{(\Gamma)}_u 
    = -16\:\Gamma^u_{vu}\int_0^\infty d\epsilon\:\text{Re}\left[d_u\tilde d_v-\tilde d_u d_v\right]\tanh\frac{\epsilon}{2k_B T}.
\end{equation}

\begin{figure}[!htb]
    \centering
    \includegraphics[]{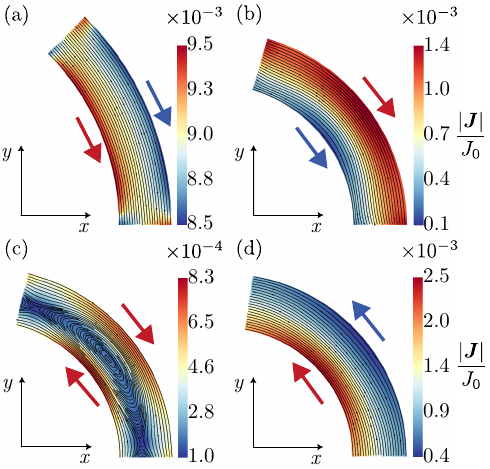} 
    \caption{\label{fig:annular_sector}Charge current density Eq.~\eqref{Eq:ChargeCurrentDensity} in an annular sector ferromagnet. The arrows indicate direction and relative magnitude on the inner and outer boundaries. Panels (a)-(d) correspond to $\kappa L/\pi=0.250$, $0.410$, $0.425$ and $0.450$, respectively.}
\end{figure}

From Eqs.~(\ref{eq:wpl_helicoid_u}-\ref{eq:current_christoffel_helicoid}), we see that the tensorial framework is particularly useful in parametrizing exchange fields that follow the geometry. For the helical strip we may take $\bm h=h_0\partial_u\mathcal{S}/|\partial_u\mathcal{S}|=h_0\bm e_u/|\bm e_u|$. This represents the ground state of a film with easy-plane anisotropy, which is expected to remain accurate for low curvatures \cite{Salamone2022,Sheka2015}. Beyond a critical curvature, the magnetic ground state will typically acquire an out-of-plane component \cite{Sheka2015}, which will give a small renormalization to the expected results but will not alter the qualitative predictions here. The triplet vector $\bm{d}$ can be decomposed into a short-ranged component defined by $d_\parallel=\bm{d}\cdot\hat{\bm{h}}$ and a long-ranged component defined by $\bm{d}_\perp=\bm{d}-d_\parallel\hat{\bm{h}}$, where the physical spins are perpendicular to the d-vector. From this we identify $d_u \sim d_\parallel$ and $d_v\sim d_\perp$ as short-range and long-range components, respectively. First derivatives in the Usadel equation correspond to spin precession, while spin relaxation enters as imaginary terms in the energy. We see that Eqs.~(\ref{eq:wpl_helicoid_u}) and (\ref{eq:wpl_helicoid_v}) couple the different triplet species, and that the spin precession driving their interconversion is proportional to the Christoffel symbols, which increase with larger curvature and torsion. Furthermore, we see that the curvature dependent current of Eq.~\eqref{eq:current_christoffel_helicoid} indeed arises due to an interconversion between short-range triplets ($d_u$) and long-range triplets ($d_v$). In the limit of a slowly varying exchange field and energy, we can see that Eqs.~(\ref{eq:current_triplet_0}) and ~(\ref{eq:current_triplet_Christoffel}) should have opposite sign for a helicoid geometry. The relative strengths of these components depend on the curvature, and when Eq.~(\ref{eq:current_triplet_Christoffel}) dominates we get a purely geometry-controlled current switching, as demonstrated in Fig.~\ref{fig:annular_sector}. More generally, we can see that we will always have non-zero Christoffel symbols when the exchange field follows the geometry and the basis vectors rotate, which will give curvature-related effects in transport observables.

\paragraph*{Numerical solution.}
When solving Eq.~\eqref{eq:Usadel} numerically, the surface geometry can either be included in the Usadel equation via a local coordinate system according to the surface parameterization, as in Eq.~\eqref{eq:Usadel_helicoid}, allowing us to use a flat geometric domain, or one can solve the equations on the physical domain in Cartesian coordinates. We saw above that interpretation of the relevant mechanisms was straight forward using the equations on the curved geometries directly. However, it is computationally favourable to find the numerical solution with the finite element method by solving the Usadel equation directly on the physical domain. The construction of such a domain relies on an appropriate mathematical representation of the surface; for an overview of surface parameterizations, see for example \cite{Krivoshapko2015}.
For numerical purposes, we express the quasiclassical retarded Green's function in the Riccati parametrization \cite{Schopohl1995, Jacobsen2015b},
\begin{equation}\label{eq:g^R_Riccati}
    \hat g^R = \begin{pmatrix}
        N(1+\gamma\tilde\gamma) & 2N\gamma \\
        -2\tilde N\tilde \gamma & -N(1+\tilde\gamma\gamma)
    \end{pmatrix},
\end{equation}
where $\gamma$ is a $2\times 2$ matrix and $N=(1-\gamma\tilde\gamma)^{-1}$. In Cartesian coordinates, the Usadel equation and boundary condition for $\gamma$ become \cite{Jacobsen2015b}
\begin{gather}
    \nabla^2 \gamma + Q(\gamma,\tilde\gamma) = 0,\label{eq:Usadel_Riccati}\\
    \bm n\cdot \nabla \gamma_1 = B(\gamma,\tilde \gamma),\label{eq:KL_BC_Riccati}
\end{gather}
where $Q=2D_F\nabla\gamma\cdot \tilde N\tilde\gamma\nabla\gamma+2i\epsilon\gamma+
    i\bm h\cdot(\bm\sigma\gamma-\gamma\bm\sigma^*)$ and $B=\frac{1}{L_1\zeta_1}(1-\gamma_1\tilde\gamma_2)N_2(\gamma_2-\gamma_1)$, with corresponding expressions for $\tilde \gamma$.
 
To solve Eq. \eqref{eq:Usadel_Riccati} with the finite element method, we use the FEniCSx Python library \cite{Baratta2023, Scroggs2022, Scroggs2022b, Alnaes2014}, which requires Eq. \eqref{eq:Usadel_Riccati} and \eqref{eq:KL_BC_Riccati} to be expressed in variational form.
This is done by multiplying Eq. \eqref{eq:Usadel_Riccati} with a $2\times 2$ matrix test function $v_\gamma$ and integrating over the geometric domain $\Omega$. The resulting equation is required to hold for all test functions $v_\gamma$ in some suitable function space $V_\gamma$. The resulting variational form becomes
\begin{equation}\label{eq:variational_form_gamma}
    \begin{aligned}
        F_\gamma(\gamma, \tilde \gamma; v_\gamma) 
        \equiv 
        &\int_\Omega dx\: \left(Q: v_\gamma-\nabla\gamma:\nabla v_\gamma\right)
        \\
        &+  
        \int_{\partial\Omega}ds\: B: v_\gamma
        =0,
        \quad \forall v_\gamma \in V_\gamma,
    \end{aligned}
\end{equation}
where we have used the divergence theorem to include the boundary condition from Eq. \eqref{eq:KL_BC_Riccati}.
Here, $\partial \Omega$ denotes the boundary of $\Omega$, $dx$ denotes the differential element for integration over $\Omega$, and $ds$ denotes the differential element for integration over $\partial \Omega$. The operator $:$ denotes the Frobenius inner product, defined for matrices $ A$ and $B$ as $A:B=\sum_{i,j} A^*_{ij}B_{ij}$. Similarly, one can derive the variational form $F_{\tilde\gamma}(\gamma,\tilde\gamma;v_{\tilde\gamma})$ for the tilde conjugated version of Eq. \eqref{eq:Usadel_Riccati}.
By defining a vector $\bm \gamma=(\gamma,\tilde\gamma)$ of unknowns and a vector $\bm v=(v_\gamma, v_{\tilde\gamma})$ of test functions in a mixed (product) function space $V=V_{\gamma} \times V_{\tilde\gamma}$, the total variational form for the system of equations can be written as
\begin{equation}\label{eq:variational_form}
    F(\bm \gamma, \bm v)\equiv F_\gamma(\gamma,\tilde\gamma;v_\gamma)+F_{\tilde\gamma}(\gamma,\tilde\gamma;v_{\tilde\gamma})=0,\quad \forall \bm v\in V.
\end{equation}
This variational formulation can then be implemented numerically together with a discretized finite element mesh of the domain.
\paragraph*{Results.}
We will consider the charge current density in a curved ferromagnetic thin film/ribbon, attached to a bulk, conventional, spin-singlet $s$-wave superconductor at either end in a superconductor-ferromagnet-superconductor Josephson junction.
In solving the Usadel equations, we fix the interface parameter $\zeta=3$, set the temperature $T=0.005 T_c$, where $T_c$ is the critical temperature of the superconductors, and we fix the phase difference between the superconductors to be $\phi=\pi/2$, which is the phase for which maximal current flows in one dimension \cite{Josephson1962}. In addition, we use the Dynes approximation to phenomenologically account for inelastic quasiparticle scattering, here with $\epsilon \rightarrow \epsilon+10^{-3}i$, and introduce an energy cutoff at $4\Delta$ in the numerical integral. For these examples, we take length $L=2\xi$ and width $W=0.5\xi$, and we set the strength of the exchange field to $h_0=\Delta$. For the finite element method, we use a second order Lagrange function space on a domain of $14\times 8$ quadrilateral elements, where the Green's function solution has stabilized.

\par The helicoidal strip is generated from a helical center line by attaching straight rulings, with a surface parametrization given by Eq.~\eqref{eq:helicoid_surface}. It reduces to an annular strip lying in the $xy$-plane when $\tau=0$. From this simple system, we can extract the dominant mechanisms driving the curvature-induced features. We solve the annular system in Fig.~\ref{fig:annular_sector}, with increasing curvatures from (a)-(d). As the curvature increases, the charge current gradually decreases before eventually reversing direction. This reversal first occurs along the inner arc, where the curvature is greatest. At $\kappa=0.425\pi/L$, shown in Fig.~\hyperref[fig:annular_sector]{1(c)}, the current along the inner arc has reversed, whereas the current along the outer arc remains unchanged. These oppositely directed currents give rise to circulating eddy currents in the middle of the strip. Although the inner and outer arcs of the annular sector differ in length, the system considered in Fig.~\ref{fig:annular_sector} lies well outside the regime where a length-induced $0-\pi$ transition occurs, as shown in Appendix~\ref{section:B}, and it is straight forward to show that such a geometry-induced transition does not occur for a curved magnet in an externally applied magnetic field of fixed direction. The observed $0-\pi$ transition here is instead caused by the effective field inhomogeneity caused by the ground state magnetization following the curve, which yields non-zero Christoffel symbols in the local frame, and results in the contribution $J_i^{(\Gamma)}$ given by Eq.~\eqref{eq:current_triplet_Christoffel}. This flows opposite to the other current contributions and grows in magnitude with increasing curvature.
\par
\begin{figure}[!htb]
    \centering
    \includegraphics[]{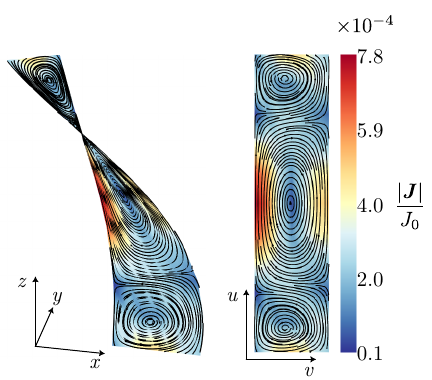} 
    \caption{Charge current density in a helicoidal ferromagnet shown in physical space (left) and in the parameter space $(u,v)$ (right). Parameters are $\kappa=0.5\pi/L$ and $\tau=0.7\pi/L$.}
    \label{fig:helicoid}
\end{figure}

 For finite $\tau$, even more complex eddy-current patterns can emerge. Fig. \ref{fig:helicoid} shows a helicoidal strip with $\kappa=0.5\pi/L$ and $\tau=0.7\pi/L$. In this case, three distinct eddy currents are present. The central eddy current circulates clockwise, while the upper and lower eddy currents circulate counterclockwise. 
These circulating currents are analogous to the spin-orbit induced supercurrent vortices found in flat films with spin orbit coupling \cite{Amundsen2017}.

\textit{Discussion.}
In many respects, experimental progress in the field of 3D curvilinear magnetism has been outpacing theoretical predictions, with several fundamental questions still unanswered, particularly regarding how it affects superconductivity. In this article, we have shown how the competing mechanisms of distance and curvature in conventional Josephson junctions with a curved ferromagnetic thin film affect the current distribution in the film. While supercurrents are expected to always travel the shortest route, we have seen that the effect of curvature in converting between singlet and triplet superconducting correlations modify the optimal pathways in such a way as to induce eddy currents, and can lead to a fully geometrically-induced current switching. We have provided a generalized theoretical Green's function framework for analyzing diffusive transport on arbitrarily curved thin film surfaces, and provided two experimentally relevant examples. Bespoke, intricate nanoscale geometries can now be manufactured in a number of different ways, for example with focused electron beam induced deposition \cite{Dobrovolskiy2021,Sanz-Hernandez2020}, atomic layer deposition \cite{Grundler2026_ald}, and glancing angle deposition \cite{Gibbs2014}, for which this diffusive transport framework is well suited, and such current-density distributions on magnetic surfaces may be measured with magneto-optical imaging \cite{Johansen1996}.

We have here treated the superconducting leads as bulk, conventional, singlet $s$-wave superconductors, thereby ignoring the reverse proximity effect of the magnet on the superconductors themselves. For larger exchange fields, accurate numerical simulation would require a straightforward extension of the numerics to include the suppression of the superconducting order via a self-consistent solution to the Usadel equation. Large curvatures are also expected to alter the magnetic ground state configuration, which should be taken account of when constructing the exchange field tensor.

The films shown here are free from bending strain, i.e. they have not been constructed to have this profile by twisting or bending a flat film. In 1D, bending strain in a ferromagnet is simply additive to the curvature-induced effect, and so amounts to a rescaling \cite{Salamone2022}. However, bending strain can introduce an \textit{additional} mechanism for controlling conversion between superconducting correlations, and may therefore be used for in-situ control of the curvature-related properties \cite{Salamone2022}. These effects can be studied by replacing the derivatives in the Usadel equation by their fully space-gauge-covariant counterparts when including the $SU(2)$ gauge potential of the spin-orbit field acting on the spin space. Similarly, a magnetic vector potential, for example from an external magnetic field or a phase gradient from a supercurrent, would enter as a $U(1)$ gauge field, this time acting on the particle-hole/Nambu space. 

It is worth noting that bending strain in the conventional s-wave superconductors, rather than in the weak link, can lead to the stabilization of spin-polarized $p$-wave superconducting correlations in clean systems, which leads to the surprising result of an induced magnetization in a Josephson junction with a normal metal \cite{Heinrich2025}. In unconventional iron chalcogenide, bending-strain appears to induce a phase transition by controlling the competition between spin fluctuation and nematic order \cite{Gao2025}. It would therefore be very interesting to extend the current curvature framework to systems that include both unconventional superconducting leads, and curvature in the entire system. In addition, one could extend the analysis to other classes of magnetic order: for example, the study of curvature effects in antiferromagnets is still in its infancy  
\cite{Pylypovskyi2020,Salamone2024}.

Recent work has shown vortex refraction and enhanced vortex velocities at tilted superconductor-normal metal interfaces \cite{Grundler2026}, as well as vortex bending in reconfigurable 3D superconducting nanostructures \cite{Zhakina2025}. It would be informative to compare how the herein demonstrated geometry-induced and length-induced (see Appendix~\ref{section:B}) mechanisms for generating circulating currents in ferromagnetic films would interact with flux-induced vortices and the resulting consequences for the vortex path and velocity.

While we have here looked at the fundamental mechanisms affecting steady-state transport of Josephson currents, many relevant technologies use externally applied currents, for example for electrical control of domain walls. While numerically intensive, extending the curvature framework to make use of the full power of the non-equilibrium Green's functions, whereby the distribution functions are calculated self-consistently in the Usadel equation \cite{Jacobsen2017}, would allow simulation of such systems. It would also be straight forward to examine spin current transport in curved systems, where, although individual spin-polarized components of the spin current would not be conserved due to the effective gradient in the exchange field, the interference between the different triplet correlations should nevertheless result in an exchange-independent spin current component \cite{Jacobsen2016}. More generally, this theoretical framework for superconducting transport in curved geometries opens up for exploring questions about holonomy in condensed matter \cite{Frustaglia2025}.

\begin{acknowledgments}
We thank M. Amundsen for insightful comments. We acknowledge funding via the ``Outstanding Academic Fellows'' program at NTNU and the Research Council of Norway Grant Nos. 262633, 354571 and 302315.
\end{acknowledgments}


%

\appendix

\section{The metric and curvature}\label{Appendix:A}

In order to explicitly calculate the metric, we utilize the formalism of the first- and second- fundamental forms \cite{Pressley2010}. This lets us write these expressions in terms of physically intuitive quantities, such as curvature. The $3$D coordinate vector is chosen to be $\bm{R}(u,v,n)=\bm{\mathcal{S}}(u,v)+n\hat{\bm{\mathcal{N}}}(u,v)$ where the surface patch $\bm{\mathcal{S}}$ describes the $2$D surface, and $\hat{\bm{\mathcal{N}}}(u,v)$ denotes the surface's normal vector. In this frame, the limit $n\rightarrow0$ confines particles to the surface $\bm{\mathcal{S}}$, corresponding to the case of a thin film material. The first fundamental form is given by $E=|\partial_u\bm{\mathcal{S}}|^2$, $F=\partial_u\bm{\mathcal{S}}\cdot\partial_v\bm{\mathcal{S}}$ and $G=|\partial_v\bm{\mathcal{S}}|^2$. This form describes distances and angles of paths on the surface, and can therefore be described as an effective $2$D metric when ordered as a matrix, 
\begin{equation}
    \mu=\begin{pmatrix} E&F\\F&G \end{pmatrix}.
\end{equation}

The second fundamental form is related to the curvature of the surface and how the surface is embedded into the surrounding space. It is natural to the define curvature in terms of how the tangential vector changes direction along the surface, or equivalently, how the normal vector changes. The second fundamental form is hence given by $\hat{\bm{\mathcal{N}}}\cdot\partial_{uu}\bm{\mathcal{S}}=L$, $\hat{\bm{\mathcal{N}}}\cdot\partial_{uv}\bm{\mathcal{S}}=M$ and $\hat{\bm{\mathcal{N}}}\cdot\partial_{vv}\bm{\mathcal{S}}=N$, which can be ordered as a matrix
\begin{equation}
    \mathcal{F} = \begin{pmatrix} L&M\\M&N \end{pmatrix}.
\end{equation}
We would like our expression for curvature to be invariant under a change of coordinates. This is not the case for the second fundamental form since this form consists of inner products between covariant vectors. This can be amended by including the inverse metric, giving
\begin{equation}
    \mathcal{W} = \mu^{-1}\mathcal{F} = \frac{1}{|\mu|}\begin{pmatrix} LG-FM&MG-FN\\ME-FL&NE-FM\end{pmatrix},
\end{equation}
where $|\mu|$ refers to the determinant of $\mu$. $\mathcal{W}$ is known as the \textit{Weintgarten matrix} or the \textit{shape operator}. There are multiple ways of defining curvature, such as principle curvatures, or mean and Gaussian curvature. The principal curvatures $\kappa_1$ and $\kappa_2$ are given by the eigenvalues of $\mathcal{W}$, and the mean and Gaussian curvatures are given by the trace and determinant of $\mathcal{W}$, respectively.

The full $3$D metric can now be computed in terms of the fundamental forms, giving us 
\begin{equation} \label{eq20}
    \mathcal{G}=\begin{pmatrix} \mathcal{G}_{11} & \mathcal{G}_{12} & 0 \\ \mathcal{G}_{21} & \mathcal{G}_{22} & 0 \\ 0 & 0 & 1\end{pmatrix},
\end{equation}
\noindent where
\begin{align}
    &\mathcal{G}_{11} =E-2nL+n^2(\mathcal{W}_{11}L+\mathcal{W}_{12}M),\\
    &\mathcal{G}_{12}=\eta_{21} =F-2nM+n^2(\mathcal{W}_{11}M+\mathcal{W}_{12}N),\\
    &\mathcal{G}_{22} =G-2nN+n^2(\mathcal{W}_{21}M+\mathcal{W}_{22}N).
\end{align}
This expression can be written more compactly, as $\mathcal{G}_{ij}=\mu_{ij}-2n\mathcal{F}_{ij} +n^2\mathcal{W}_{ik}\mathcal{F}_{kj}$. In the thin film limit where $n\rightarrow0$, $\mathcal{G}_{ij}$ reduces to $\mu_{ij}$, as expected.

Taking the derivative of vector quantities produce additional terms, $\partial_i\bm{A}= (\partial_iA_j)\bm{e}^j + A_j(\partial_i\bm{e}^j)$. The derivatives of basis vectors are usually expressed in terms of Christoffel symbols, $\Gamma^k_{ij}$, which can be found from either $\partial_i\bm{e}^j=-\Gamma^j_{ik}\bm{e}^k$, $\partial_i\bm{e}_j=\Gamma^k_{ij}\bm{e}_k$ or $\Gamma^k_{ij}=\frac{1}{2}\mathcal{G}^{kl}\left(\partial_j\mathcal{G}_{li}+\partial_i\mathcal{G}_{lj}-\partial_l\mathcal{G}_{ij}\right)$, such that we can write
\begin{equation} \label{eq:gauge_cov_derivative}
    \partial_i\bm{A}=(\partial_iA_j - \Gamma^k_{ij}A_k)\bm{e}^j.
\end{equation}
In the thin film limit, the non-zero Christoffel symbols can now be shown to be given by \cite{Pressley2010}
\begin{align}
    \Gamma^1_{11}&=\frac{1}{2|g|}(GE_u -2FF_u + FE_v),\\
    \Gamma^1_{12}&=\frac{1}{2|g|}(GE_v -FG_u), \\
    \Gamma^2_{11}& =\frac{1}{2|g|}(2EF_u -EE_v-FE_u), \\
    \Gamma^2_{12}& =\frac{1}{2|g|}(EG_u -FE_v), \\
    \Gamma^1_{22}&=\frac{1}{2|g|}(2GF_v -GG_u-FG_v), \\
    \Gamma^2_{22} &=\frac{1}{2|g|}(EG_v -2FF_v+FG_u), \\
    \Gamma^1_{13}&=-\mathcal{W}_{11}, \quad \Gamma^1_{23} =-\mathcal{W}_{12}, \label{eq:christoffel_symbol_n_dir_1} \\\Gamma^2_{13} &=-\mathcal{W}_{21}, \quad \Gamma^2_{23} =-\mathcal{W}_{22}. \label{eq:christoffel_symbol_n_dir_2}
\end{align}
Although the Christoffel symbols of Eqs.~(\ref{eq:christoffel_symbol_n_dir_1},\ref{eq:christoffel_symbol_n_dir_2}) contain normal-direction indices, they are useful in the thin film limit for evaluating $3$D vectors on the surface, such as magnetization.

\section{The weak proximity limit} \label{Appendix:WPL}

In order to draw analytical insight from the Usadel equation it is standard to consider the weak proximity limit, where the proximity effect is weak, such that we can simplify the retarded Green's function as
\begin{equation}
    \hat{g}^R=\left(\begin{matrix} 1 & f \\-\Tilde{f} & -1 \end{matrix}\right),
\end{equation}
where the anomalous Green's function is considered to be small, such that $f\ll1$. The Usadel equation in this limit can be found from Eq.~\eqref{eq:Usadel_Riccati} by setting $\gamma=f/2$ and $N=1$, and linearizing to first order in $f$, giving
\begin{equation}
     \mathcal{G}^{ij}D_F\partial_i\partial_jf-\Gamma^k_{ij}\partial_kf = -2iEf-i\mathcal{G}^{ij}h_i(\sigma_jf-f\sigma_j^*),
\end{equation}
which is the linearized Usadel equation.
Here, we have used Eq.~\eqref{eq:gauge_cov_derivative}, giving
\begin{equation*}
    \nabla^2 f=\mathcal{G}^{ij}\left(\partial_i\partial_j f - \Gamma^k_{ij}\partial_kf\right).
\end{equation*}

Writing $f$ in the spin-parameterization, $f=(f^{(s)}+d^\alpha \sigma_\alpha)i\sigma_y$, gives
\begin{equation} \label{wpl_2D_d}
    \begin{split}
        &D_F\mathcal{G}^{ij}(\partial_i\partial_jd_\alpha - 2\Gamma^\beta_{j\alpha}\partial_id_\beta- \Gamma^k_{ij}\partial_k d_\alpha \\&- d_\beta(\partial_i\Gamma^\beta_{j\alpha} - \Gamma^\beta_{i\gamma}\Gamma^\gamma_{j\alpha}- \Gamma^k_{ij}\Gamma^\beta_{k\alpha})) =-2i(Ed_\alpha+f^{(s)}h_\alpha),
    \end{split}
\end{equation}
which corresponds to one equation of motion for each of the three $d$-vector components by letting $\alpha=u$, $v$, $n$. We note that the sums in~\eqref{wpl_2D_d} should be taken such that $i,j,k=u,v$ and $\beta=u,v,n$ ensuring that motion is constrained to the $2$D surface $\bm{\mathcal{S}}(u,v)$, while the $3$D vector $\bm{d}$ may point in all $3$ directions. For the singlet component, we get
\begin{equation} \label{wpl_2D_f}
    D_Fg^{ij}(\partial_i\partial_jf^{(s)}-\Gamma^k_{ij}\partial_kf^{(s)}) =-2i(Ef^{(s)}+\mathcal{G}^{\alpha\beta}h_\alpha d_\beta).
\end{equation}
The triplet vector $\bm{d}$ can be decomposed into a short-ranged part defined by $d_\parallel=\bm{d}\cdot\hat{\bm{h}}$ and a long-ranged part defined by $\bm{d}_\perp=\bm{d}-d_\parallel\hat{\bm{h}}$. Generally, the exchange field $\bm{h}$ tends to align with the curvature, i.e. $\bm h=h_0\bm e_u/|\bm e_u|$, from which we can see that $d_\parallel\sim d_u$. In such cases, we can see from Eq.~\eqref{wpl_2D_d} that conversion between long-range triplets and short-range triplets is indeed facilitated by the geometric curvature represented by the Christoffel symbols, as expected. 

\section{Length effects}\label{section:B}
\renewcommand{\thefigure}{B\arabic{figure}}
\setcounter{figure}{0}

\begin{figure}[!ht]
    \centering
    \includegraphics[]{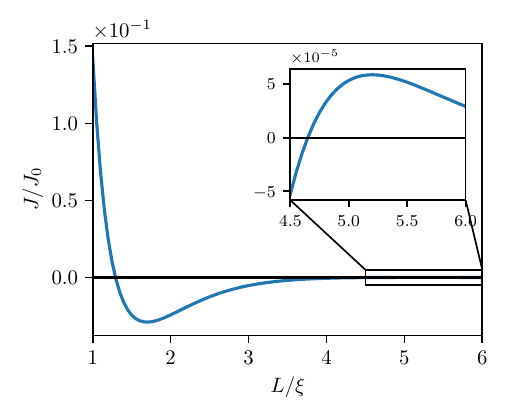} 
    \caption{Charge current density [Eq.~\eqref{Eq:ChargeCurrentDensity}] at a single point in a rectangular ferromagnet, with $W=0.5\xi$ and $h_0=\Delta$ in the longitudinal direction, for different values of the length $L$. Inset shows the second crossover.} 
    \label{fig:appendix_length}
\end{figure}

\begin{figure}[!htb]
    \centering
    \includegraphics[]{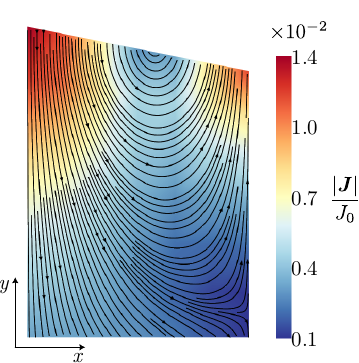} 
    \caption{Charge current density in a trapezoidal ferromagnet. Parameters are $W=1.0\xi$, left length $L_1=1.4\xi$, right length $L_2=1.2\xi$ and $\bm h=\Delta \hat{\bm y}$.}
    \label{fig:appendix_trapezoid}
\end{figure}

To illustrate where the length-induced charge current reversals occur in an SFS junction near these junction parameters, we calculate the charge current density Eq.~\eqref{Eq:ChargeCurrentDensity} for a rectangular ferromagnet with $W=0.5\xi$ and $h_0=\Delta$ for different values of $L$, with the magnetic field oriented along the longitudinal direction. For such a system, the current density is uniform throughout the ferromagnet, so it suffices to consider its value at a single point. The results shown in Fig.~\ref{fig:appendix_length} exhibit $0-\pi$ transitions at $L\approx1.3\xi$ and $L\approx 4.6\xi$.
This results in length-driven current reversals and can lead to length-induced circulating currents, as shown here for a trapezoidal ferromagnet in Fig.~\ref{fig:appendix_trapezoid}. The left and right edges have lengths $L_1=1.4\xi$ and $L_2=1.2\xi$, respectively, placing them on opposite sides of the first transition at $L=1.3\xi$. Consequently, the current flows in opposite directions on either side of the ferromagnet. Fig.~\ref{fig:appendix_length} also shows that the eddy currents observed for the annular sector in Fig.~\ref{fig:annular_sector} and the helicoid in Fig.~\ref{fig:helicoid} are not driven by this length effect, as their central length of $L=2\xi$ is well outside the range where the length-induced $0-\pi$ transition occurs.

\end{document}